\documentclass[twocolumn,prl,showpacs,amsmath,amssymb]{revtex4}
\usepackage{graphics,graphpap,amssymb,psfrag,subfigure,picinpar}
\usepackage{graphicx}
\usepackage{color}
\begin{document}

\title{Superconductor insulator transition in thin films driven by an orbital parallel magnetic field effect}

\author{{\sc Dganit Meidan$^1$ and Yuval Oreg$^{1,2}$}\\
{\small \em $^1$Department of Condensed Matter Physics, Weizmann
Institute of Science, Rehovot, 76100, ISRAEL\\
$^2$Applied Physics Department, Stanford University, Stanford, California 94305-4090, USA}}

\begin{abstract}
We study theoretically orbital effects of a parallel magnetic field applied to a disordered superconducting film. We find that the field reduces the phase stiffness and leads to strong quantum
phase fluctuations driving the system into an insulating behavior. This microscopic model shows that the
critical field decreases with the sheet resistance, in agreement
with recent experimental results. The predictions of this model can be used to discriminate spin and orbital effects. We find that experiments conducted by A.~Johansson \textit{et al.} are more consistent with the orbital mechanism.
\end{abstract}

\maketitle

\emph{Introduction and results }- Applying a magnetic field to a disordered superconducting~(SC)
film, can drive it into a strong insulating~(INS) state. This was
observed 
both when the film is placed perpendicular to the field, in
InO~\cite{AFHebardPRL1990,VFGantmakherJETPL2000,GSambandamurthyPRL2004},
MoGe~\cite{AYazdaniPRL1995}, TiN~\cite{TIBaturinaJETPL2004} and
NbSe~\cite{HAubinPRB2006}, or when the film is in parallel to the
field orientation in
InO~\cite{VFGantmakherJETPL2000_2,AJohanssonCondMat2006} and
Bi~\cite{KAParendoPRB2006}. While several theoretical models
support a SC-INS transition in the perpendicular
orientation~\cite{MPAFisherPRL1990,YDubiNAT2007}, the mechanism
that drives the transition when the field is parallel to the film,
remains unclear.

We study analytically a microscopic model of a
disordered SC film with magnetic field applied \emph{parallel} to
the film, focusing on the induced orbital effects. We show that
the field reduces the stiffness of the SC phase, leading to strong quantum
phase fluctuations manifested as an INS behavior. We find that the
transition to the INS phase occurs at a critical field that
depends on the SC coherence length, $\xi_0 $, the film's
thickness, $d$, and the sheet resistance, $R_\square $ [see Eq.
(\ref{critical_parallel_field_vs_resistance})]. We will show that
this relation does not depend on the detailed mechanism that drives the transition,
that it allows to determine experimentally if spin or orbital
effects are dominant in parallel field, and that the measurements of A.~Johansson \emph{et al.}~\cite{AJohanssonCondMat2006} are more consistent with the orbital mechanism.

The emergence of a SC-INS transition in disordered films induced by a
\emph{perpendicular} magnetic field is consistent with several
theoretical scenarios. A \emph{perpendicular} magnetic field penetrates the film in the form of vortices. As the field increases, these vortices were predicted to delocalize and Bose-condense leading to an INS behavior~\cite{MPAFisherPRL1990}. An alternative numerical work studied the effect of thermal phase fluctuations in a disordered 2D SC~\cite{YDubiNAT2007}. The \emph{perpendicular} magnetic field was shown to destroy phase correlations between SC islands. Conversely, a \emph{parallel} spin-exchange field causes the order parameter phase and amplitude to vanish abruptly. While existing theories can account for the qualitative behavior seen in the \emph{perpendicular} field orientation, they do not explain the surprisingly similar observed phenomenology when the field is \emph{parallel} to the film~\cite{VFGantmakherJETPL2000_2,AJohanssonCondMat2006,KAParendoPRB2006}.

\begin{figure}[h]
\begin{center}
\includegraphics[width=0.5\textwidth]{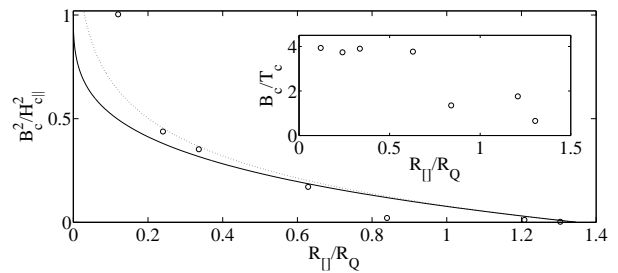}
\caption[width=0.5\textwidth]{Circles indicate the measured $B_c^2 $ in units of
$H_{c\parallel}^2=(1.6)^2\phi_0^2 3  T_c\nu_0/(\gamma d)R_\square/R_Q$ versus the normalized resistance per square, $R_\square/R_Q $, from Ref. \onlinecite{AJohanssonCondMat2006}. Here $T_c$ is taken from an
independent measurement. The solid and dashed lines mark a fit of Eq. (\ref{critical_parallel_field_vs_resistance}) to the
data, for $\Delta(B_c) $ and $\Delta(B=0) $, respectively, using $K_0^c$ and $\nu_0$ as fitting parameters. $K_0^c$ determines the maximal resistance films that exhibit a SC phase. The fitting values are $K_0^c=0.37$, and $\nu_0 \sim 4.8 \times 10^{33}\textrm{erg}^{-1}\textrm{cm}^{-3}$. Estimates
for $K_0^c$ based on the proliferation of vortex loops give $ 0.46<K_0^c<0.75$, see
Fig. \ref{fig:critical stiffness}. $\nu_0 $ inferred from typical
carrier densities in amorphous InO is $\nu_0 \approx 1.4
\times 10^{33}\textrm{erg}^{-1}\textrm{cm}^{-1}$. Inset: The measured $B_c/T_c $, versus $R_\square/R_Q $. A transition driven by spin-exchange effects would imply that the ratio $B_c/ T_c $ be independent of $R_\square $. Here, we find a large scattering of the data.
  Moreover, in the presence of spin-orbit scattering, increasing the disorder (large $R_\square $) would enhance spin-orbit scattering, resulting in a larger $B_c$, whereas the data of Ref.~\onlinecite{AJohanssonCondMat2006} shows the opposite tendency.} \label{fig:parallel_fit_shahar}
\end{center}
\end{figure}

We study the previously disregarded orbital effect of a \emph{parallel} magnetic
field applied to a disordered SC film. We find that the field uniformly decreases the SC order parameter and reduces its phase stiffness. As a result, quantum fluctuations of the phase and amplitude are
enhanced and can lead to an INS behavior. Our main prediction is
that the critical field,~$B_c$ that marks the onset of
the INS behavior depends on the critical temperature~$T_c$, the
film's sheet resistance,~$R_\square$ and thickness~$d$ as
\begin{eqnarray}\label{critical_parallel_field_vs_resistance}
 B_c^2/\tilde{H}^2&=&1/2\left\{\ln\left(R_Q/R_\square\right)-\ln{2K_0^c}\right\},
\end{eqnarray}
where $\tilde{H}^2 = \frac{12T_c\phi_0^2\nu_0 }{\pi\gamma
d}\frac{\Delta_0(B)}{\Delta_0}\frac{R_\square}{R_Q}$, $\gamma=1.78 $, $\phi_0=hc/2e$ is the flux quantum, $\nu_0 $ is the density of states, $R_Q = h/(4e^2) $ is the resistance quantum, $\Delta_0(B)$ is the mean field order parameter in the presence of a pair breaking field \cite{KMaki1969}, and $K_0^c$ is the critical value of the stiffness coefficient. Three points should be stressed herein. First, while the INS behavior can be a result of
either the proliferation of topological phase
excitations, or
of strong gaussian phase fluctuations, the detailed
mechanism that drives the transition will merely change the
numerical factor $K_0^c$, as long as dissipation in the cores is negligible~\cite{CommentDissipationInCores,RBundschuhNPB1998}. Second, we do not consider the effects of a spin-exchange field, as some of the materials that exhibit
the SC-INS transition, such as MoGe, NbSe and Bi, are expected to
have strong spin-orbit scattering, which would smear spin polarization effects. However, while the spin mechanism is
expected to depend on the thickness, $d$, only implicitly through $T_c$, the orbital critical filed is inversely
proportional to $d$. Hence, by studying the dependance of $B_c $ field on the parameters of the system such as $R_\square$, $d $ and $T_c $, one can determine which mechanism dominates the transition. Finally, we find that the data of Ref. \onlinecite{AJohanssonCondMat2006} is more consistent with Eq.~(\ref{critical_parallel_field_vs_resistance}), see Fig.~\ref{fig:parallel_fit_shahar}.

To gain insight to how the reduced stiffness can lead to
an INS behavior we consider as an example its effect on quantum vortex
excitations, corresponding to a vortex loop in the $D=2+1$
dimensional system. The energy cost of a circular vortex loop is $E_{\textrm{loop}}
\sim E_{\textrm{core}} L/\xi + E_{\textrm{J}} L/\xi \ln{L/\xi}$,
where $E_{\textrm{core}}$ is the energy per unit length to destroy
the superconductor in the core of the vortex and $ E_{\textrm{J}}
$ is the energy per unit length to rotate the SC phase, and is
determined by the phase stiffness. The logarithmic divergence will
be cutoff for more complicated loop shapes. The entropy of a
loop is determined by counting all possible configurations. On
a cubic lattice this can be estimated by
$\texttt{S}_{\textrm{loop}}\sim\ln(2D-1)^{L/\xi}\sim L/\xi $. A
parallel magnetic field will reduce the core and rotational
energies of the vortex loops [see discussion following Eq. (\ref{uniform_action_with_B}) and (\ref{stiffness coefficient})], leaving the
entropy unchanged. As a result, vortex excitation become
increasingly favorable and proliferate at a critical field which, in the absence of dissipation inside the vortex cores~\cite{CommentDissipationInCores,RBundschuhNPB1998}
marks the onset of the INS behavior. In addition to its effects on
topological excitations, the reduced phase stiffness also enhances
gaussian fluctuation of the SC phase. Loss of phase rigidity due
to strong gaussian fluctuations give similar estimates for the
transition field as in Eq.
(\ref{critical_parallel_field_vs_resistance}), differing by a
numerical factor~$K_0^c $.

Fig. \ref{fig:parallel_fit_shahar} compares $B_c $ calculated using
Eq.(\ref{critical_parallel_field_vs_resistance}) with the
data of Ref. \onlinecite{AJohanssonCondMat2006}. The theoretical
curve was plotted with $K_0^c$ and $\nu_0$ used as fitting parameters. The microscopic model used to obtain Eq. (\ref{critical_parallel_field_vs_resistance}) is valid for $B_c/H_{c\parallel}<1$. The clean sample of Ref. \cite{AJohanssonCondMat2006} exhibit a transition into a metal at high magnetic fields. We identify $B_c$ in the cleanest sample with $H_{c\parallel}$. While our theory is not applicable for the clean sample, we use this point to determine the relation between $H_{c\parallel} $, $T_c $ and $R_\square $. The resulting $H_{c\parallel}=1.6\phi_0\sqrt{3\frac{ T_c\nu_0 }{\gamma d}\frac{R_\square}{R_Q}} $ is larger than the critical field calculated in Ref. \onlinecite{KMaki1969} by a factor $1.6$. This may be a result of the finite thickness of the films used in Ref.~\onlinecite{AJohanssonCondMat2006}. The inset of Fig. \ref{fig:parallel_fit_shahar} shows the scaling of the measured $B_c $ with the critical temperature $T_c$, as expected for a spin mechanism. We find that the data of Ref. \onlinecite{AJohanssonCondMat2006} is more consistent with Eq.~(\ref{critical_parallel_field_vs_resistance}), see caption of Fig.~\ref{fig:parallel_fit_shahar}.
We have tried to fit the experimental data of Ref. \onlinecite{KAParendoPRB2006} to our model. $B_c $ obtained using Eq. (\ref{critical_parallel_field_vs_resistance}) is in better agreement with the experimental data than the critical field obtained from a linear $B_c\propto T_c $ or a square root dependence $B_c\propto \sqrt{T_c }$, as naively expected from a spin mechanism, with weak or strong spin-orbit scattering, respectively. However, as apposed to InO \cite{AJohanssonCondMat2006}, the distinction between the two mechanism in Bi \cite{KAParendoPRB2006} is quantitative rather than qualitative.

\emph{The model} - To study the orbital effects of a parallel magnetic field on the fluctuations of the SC, we consider the microscopic action for a quasi-2D SC film, obtained from
the BCS Hamiltonian by a Hubbard-Stratanovich transformation
followed by an expansion around the saddle point
\cite{CommentDirtyLimit}. For
$T\ll\omega,Dq^2\ll\Delta_0$, this yields
\cite{UEckernJLTP1988,RASmithPRB1995,AvanOtterloCondMat1997,DMeidanPhysicaC2008}:
\begin{eqnarray}\label{effective_action_const VC}
\nonumber
&\mathcal{S}& = \nu_0d\Delta_0^2\int dxdyd\tau\left\{ \frac{\rho(x,\tau)^2}{2}\left[\ln\rho(x,\tau)^2-1\right]\right.\\
\!\!\!\!&+&\!\!\!\xi_0^2\!\!\left[\left(\nabla\rho\right)^2\!+\!\!\left(\!\frac{\partial_\tau\rho}{v_{\rho}}\!\right)^{\!\!2}
\right]
\left.+2\xi_0^2\rho^2\!\!\left[\left(\nabla\phi\right)^2+\left(\!\frac{\partial_\tau\phi}{v_{\phi}}\!\right)^{\!\!2}
\right]\!\right\}\!,
\end{eqnarray}
where $d$ is the film thickness,
$v_\rho=\sqrt{(3\pi/2)D\Delta_0}$ the amplitude velocity,
$v_\phi=\sqrt{\pi D\Delta_0(2dV_c\nu_0+1)} $ the phase velocity, $\xi_0^2=\pi D/8\Delta_0 $,
$V_c \approx  2 \pi e^2 d$ is the Fourier transform of the short range Coulomb interaction due to external screening,
$\nu_0$ the density of states, $D$ the electronic
diffusive constant, and the SC order parameter
is parameterized as $\Delta = \Delta_0\rho e^{i\phi}$, with
$\Delta_0$, the mean field solution.

A parallel magnetic field alters the mean field solution, $\Delta_0\rightarrow\Delta_0(B) $ \cite{KMaki1969}, and
couples to the gradients of the phase
$\nabla\phi\rightarrow \nabla\phi -\frac{2 e}{c}A$. We ignore
dynamic fluctuations of the electromagnetic field. This
corresponds to assuming an infinite penetration depth of the
magnetic field, and valid in the limit of very thin films. In the
London gauge the uniform action is
\begin{eqnarray}\label{uniform_action_with_B}
\!\mathcal{S} \! \!&=&\! \! \nu_0d\Delta_0^2(B)\!\!\int\!\! d^2rd\tau
\left\{\!\rho^2/2\left[\ln\rho^2-1\right]+
\rho^2B^2/\tilde{H}^2\!\right\}\!\!,
\end{eqnarray}
where $\tilde{H}^2 =\frac{12\Delta_0(B)}{\pi D e^2 d^2} $.
Taking $\rho=\alpha\rho'$, and choosing $\alpha $ to eliminate the third term in Eq.~(\ref{uniform_action_with_B}), we obtain Eq.~(\ref{effective_action_const VC}) with
$\Delta_0(B)\rightarrow \tilde{\Delta}(B)=\Delta_0(B)
\exp{\{-B^2/\tilde{H}^2\}}$, and
$\rho\rightarrow \rho'$. Minimizing the uniform action with
respect to $\Delta=\Delta_0(B)\rho'$, yields $\Delta_{MF}=\tilde{\Delta}(B)$~\cite{DMeidanPhysicaC2008}.

Motivated by existing theoretical models that stress the role of
phase fluctuations as the cause of the INS behavior~\cite{MPAFisherPRL1990,YDubiNAT2007}, we concentrate on the phase
action in Eq. (\ref{effective_action_const VC}). Taking $r\rightarrow r/(\sqrt{ D \Delta_0(B)}) $,
and rescaling the imaginary time $z = \tau\Delta_0(B) $, Eq.~(\ref{effective_action_const VC}) becomes
\begin{eqnarray}\label{microscopic_3D_XY_action}
\mathcal{S}[\phi] &=& \frac{K_0}{2}\int d^2r
dz\left((\nabla\phi)^2+\frac{(\partial_z\phi)^2}{N_\perp^2}\right),
\end{eqnarray}
where $N_\perp= p_F d$ is the number of transverse channels,
\begin{eqnarray}\label{stiffness coefficient}
  K_0 &=& \frac{\pi \nu_0 d D}{2}e^{-\frac{\pi D e^2 B^2d^2}{6\Delta_0(B)}}=\frac{R_Q}{2R_\square}e^{-2B^2/\tilde{H}^2},
\end{eqnarray}
is the stiffness coefficient. The stiffness coefficient determines the action of twisting the phase of the order parameter. When $K_0\gg 1$, the SC phase is rigid. When $K_0\ll1 $ the phase is strongly fluctuating.
Hence, there exists a critical value $ K_0^c$ that marks the onset of strong phase fluctuations, whose exact numerical value depends on the details of the transition. From Eq.~(\ref{stiffness coefficient}), this implies a critical parallel magnetic field, given by Eq.~(\ref{critical_parallel_field_vs_resistance}).

\emph{Estimate of $K_0^c $} - While the critical field in Eq.
(\ref{critical_parallel_field_vs_resistance}) depends on the
details of the transition only through $K_0^c$, this numerical
factor becomes increasingly important in the limit
$R_\square\rightarrow R_Q$. Eq. (\ref{critical_parallel_field_vs_resistance}) shows that $K_0^c$ determines the limiting value of $R_\square/R_Q$ for which $B_c\rightarrow 0$, and the SC phase disappears. To estimate the value of $K_0^c$ we
study the microscopic phase action Eq.
(\ref{microscopic_3D_XY_action}). This quantum action is in the
universality class of the anisotropic 3D XY model. The system
described by the 3D XY model undergoes a transition between an
ordered phase (SC) and a strongly fluctuating phase (INS). Different mechanisms can drive the system into a strongly fluctuating phase, including strong gaussian fluctuations and the proliferation of topological excitations. Estimates based on the Lindeman criterion give $K_0^c $ up to a numerical factor which is usually determined experimentally. A more accurate estimate can be done based on the proliferation of vortex loops, using a 3D generalization of the Kosterlitz Thouless scaling procedure~\cite{DRNelsonPRB1977,GAWilliamsPRL1987,SRShenoyPRB1989,SRShenoyPRB1995}.
We note that the critical exponents inferred from the 3D XY model
are consistent with few experimental
results~\cite{KAParendoPRB2006}.

Previous works calculated the critical stiffness for a
phase only 3D XY model, Eq. (\ref{microscopic_3D_XY_action})~\cite{SRShenoyPRB1989,SRShenoyPRB1995}. The resulting $K_0^c $
implies that films with $R_\square/R_Q>0.64 $ should not exhibit a transition, since $ B_c=0 $. Conversely, the data of Ref.~\onlinecite{AJohanssonCondMat2006} shows that a transition persists for  $R_\square/R_Q\leq1.31 $. Moreover, the phase only model
is applicable to inhomogeneous systems such as Josephson
junction arrays. In a homogenous system such as a SC film,
however, a vortex excitation can only occur once the SC amplitude
is locally suppressed. We solve the flow
Eq. (\ref{quasi_2D_scaling_eq}) and (\ref{isotropic_scaling_eq}), with corrected initial conditions to
account for both the phase rotation and the amplitude suppression
inside the vortex core, see the discussion following Eq.(\ref{isotropic_scaling_eq}). The critical bare physical parameters are the initial conditions that flow to the critical point of Eq. (\ref{isotropic_scaling_eq}). The microscopic action [Eq.~(\ref{effective_action_const VC})] allows to express
these critical parameters in terms of measurable quantities,
such as $R_\square$ and $B$.

\begin{figure}[h]
\begin{center}
\includegraphics[width=0.5\textwidth]{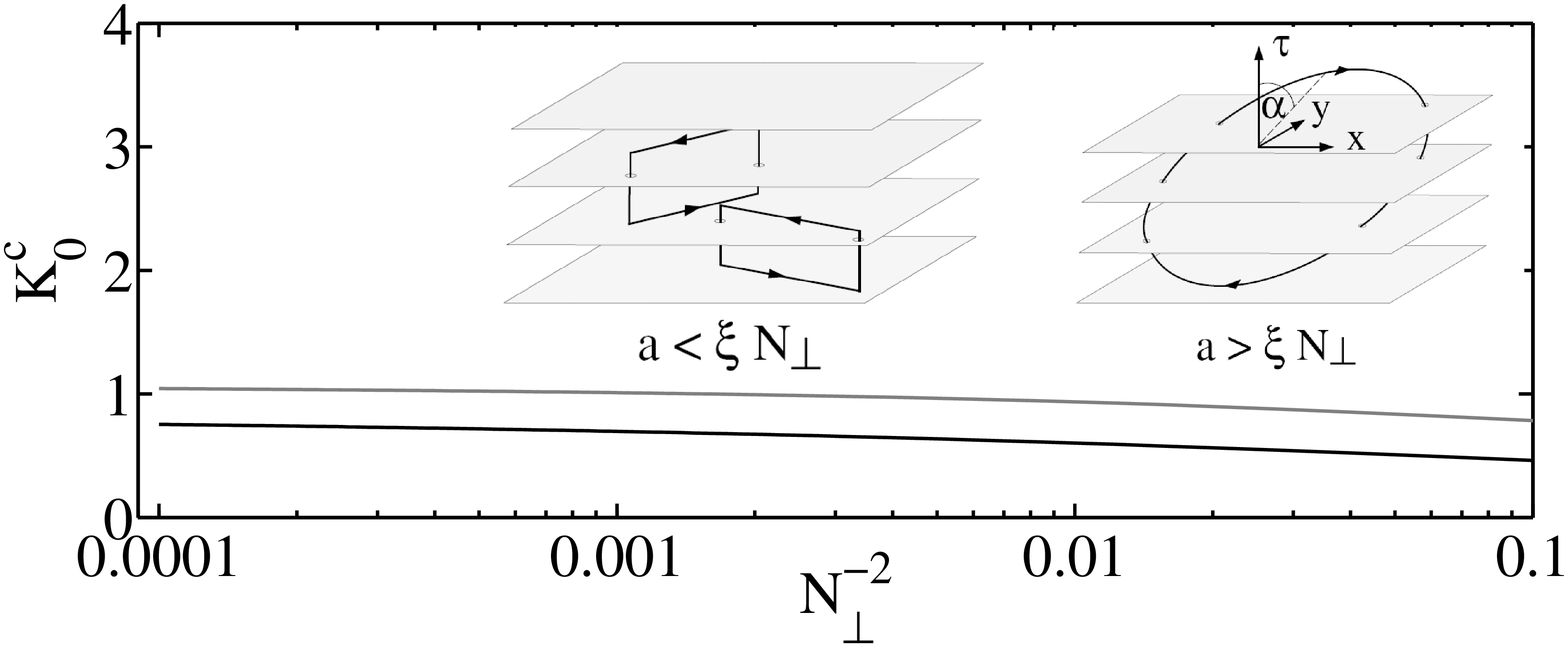}
\caption[width=0.5\textwidth]{The critical stiffness, $K_0^c$, versus the
anisotropy parameter, $N_\perp^{-2}=\frac{1}{(p_Fd)^2}$,
calculated using Eqs.~(\ref{quasi_2D_scaling_eq}) and~(\ref{isotropic_scaling_eq}), with the corrected initial conditions including both amplitude and phase fluctuations (black curve) and the phase only initial conditions (gray curve). The critical bare stiffness was
calculated assuming quasi-2D rectangular loops
\cite{SHikamiPTP1980} at scales $a<\xi N_\perp $, and multiplane
loops at scales $a>\xi N_\perp $, see inset. We find that $K_0^c$ derived for the phase and amplitude initial conditions is more consistent with the experimental data of Ref. \onlinecite{AJohanssonCondMat2006}, see discussion following Eq.(\ref{isotropic_scaling_eq}). } \label{fig:critical stiffness}
\end{center}
\end{figure}

The large number of perpendicular channels, $ N_\perp \gg1$, in quasi-2D films, generates a strong anisotropy between the spatial dimension and the imaginary time dimension, Eq. (\ref{microscopic_3D_XY_action}). This strong anisotropy in the stiffness coefficient introduces crossover scale $a'=\xi_0 N_\perp$. The scaling of the phase stiffness and vortex loop fugacity in the 3D XY model can be obtained in two regimes. At small distances $a<a'$, dominant excitations are found to be rectangular loops, cutting single
planes (left inset of Fig. \ref{fig:critical
stiffness})~\cite{SHikamiPTP1980}. The scaling equations at these scales are quasi-2D~\cite{SRShenoyPRB1995}
\begin{eqnarray}\label{quasi_2D_scaling_eq}
\nonumber
  dK_l^{2D}/dl &=& -4\pi^3 K_l^{2D}y_l^{2D}\\
  dy_l^{2D}/dl &\approx& \left[4- 2\pi
  K_l^{2D}\left(1+U(0)N_\perp^{-2}e^l/2\right)\right]y_l^{2D}\!\!,
\end{eqnarray}
where $l =\ln a$ is the running scale, and $K_l^{2D}$ and $y_l^{2D}$ are
the quasi-2D renormalized stiffness and fugacity, respectively. Here $U(0)=\sum_{q,\omega}4\pi /(q^2+ \omega^2/N_\perp^2)$ is the phase propagator, and the sum is cutoff at the effective core size that accounts for the crinkling of the vortex loops~\cite{SRShenoyPRB1995}. At larger distances $a>a'$ the system is no longer sensitive to the anisotropy, and dominant
excitations are multiplane vortex loops (right inset of Fig.~\ref{fig:critical stiffness}). In
this regime, the renormalized $K_{l'}^{2D}$ and $y_{l'}^{2D} $ of Eq.~(\ref{quasi_2D_scaling_eq}) at $l' =
\ln\left(a'/a\right)$ are used as initial conditions for the isotropic scaling equations for multiplane loops~\cite{SRShenoyPRB1989}
\begin{eqnarray}\label{isotropic_scaling_eq}
\nonumber
  dK_l/dl &=& K_l-4\pi^3/3 y_lK_l^2\\
  dy_l/dl &=& \left[6-\pi^2 K_l\left(1-x\ln K_l\right)\right]y_l.
\end{eqnarray}
 Here $x=0.6$ is the self avoiding random walk exponent. It accounts for partial cancelation of the
Biot-Savart-like interaction in complicated loop geometries
\cite{SRShenoyPRB1989,BChattopadhyayPRB1993}.

We calculate the initial conditions of
Eq.~(\ref{quasi_2D_scaling_eq}) for a homogenous system, with both phase and amplitude fluctuations.
The bare value of the stiffness coefficient is $K_0^{2D}=K_0/[1+(2 K_0)^{-1}]$~\cite{SRShenoyPRB1995}, with $K_0$ given by
Eq.~(\ref{stiffness coefficient}). The bare fugacity of a vortex loop is $y_0^{2D} = \exp{\{-\mathcal{S}_j-\mathcal{S}_c\}} $, where $ \mathcal{S}_j$ and $\mathcal{S}_c$ are the a self rotational and core actions, respectively.
Previous works calculated $\mathcal{S}_j$ for a phase only model~\cite{SRShenoyPRB1995}. The self rotational action of the smallest rectangular loop was found to be $\mathcal{S}_j=\pi^2 K_0/[1+(2K_0)^{-1}]$.
Here we add an estimate for $\mathcal{S}_c$, calculated from
Eq.~(\ref{effective_action_const VC}), using a variational method.
The anisotropy between the spatial dimension and the imaginary
time dimension introduces two possible excitations: a vortex,
manifested as a rotation of the phase in the $x-y$ plane and a
phase slip which is the corresponding rotation in the $x-\tau$ or
$y-\tau$ planes. A vortex loop in the $2+1$ dimensional world is a
complicated combination of vortex and phase slips segments. We
first estimate the core action of a vortex/phase slip segment of
unit volume, using the core size
$r_0$ as a variation parameter following Ref.~\onlinecite{DMeidanPhysicaC2008}.
The resulting core action segments are $\mathcal{S}_c^{\textrm{ps}}= K_0[\sqrt{\pi/6}+\pi/(2N_\perp)]$ and $\mathcal{S}_c^{\textrm{v}} =
 K_0\left(1+3\pi\right)/4$.
In quasi-2D films with $N_\perp\gg 1$, rectangular loops dominate~\cite{SHikamiPTP1980}. The smallest rectangular loop has $\xi_0 $ unit sides of circulation segments in the $\tau $ direction and $2\xi_0$ sides of circulation pointing in the $x-y$ plane. As a result, the bare core action of a rectangular loop is $ \mathcal{S}_c = 2 \mathcal{S}_c^{\textrm{v}}+4
\mathcal{S}_c^{\textrm{ps}}$.

The bare critical stiffness, $K_0^c(N_\perp)$, identified as the
initial condition derived for the phase and amplitude model [Eq. (\ref{effective_action_const VC})] that flow to the critical point of Eq.~(\ref{isotropic_scaling_eq}), is
plotted in black in Fig. \ref{fig:critical stiffness}. The same quantity calculated for a phase only model is plotted in gray. The corrected initial conditions give $0.46<K_0^c(N_\perp)<0.75 $, for $10<N_\perp^{2}<10^4 $. This implies that for the physically realized values of $N_\perp$ a transition could occur up to $R_\square/R_Q \leq1.08$, compared to $R_\square/R_Q \leq0.64$ found for the phase only model. Hence, the estimates for $K_0^c(N_\perp)$ based on the phase and amplitude model are more consistent with the data of Ref.~\onlinecite{AJohanssonCondMat2006}, that show a transition in samples with $R_\square/R_Q \leq 1.31$. Our estimates of $\mathcal{S}_c$ neglected the possible crinkling of the vortex loops. These would result in a larger core action, and therefore a larger range of $R_\square/R_Q $ that exhibit a transition.

\emph{Summary} - We study the previously disregarded orbital effect of a magnetic field
applied parallel to a SC film. We find that the parallel field
reduces the phase stiffness and leads to strong phase fluctuation
at a critical magnetic field that depends on the film's sheet
resistance, Eq.~(\ref{critical_parallel_field_vs_resistance}).
This prediction does not depend on the details of the transition,
it allows to experimentally determine if spin or orbital effects
drive the transition in the parallel orientation, and it shows that the data of Ref.~\onlinecite{AJohanssonCondMat2006} are more consistent with the orbital mechanism, see Fig. \ref{fig:parallel_fit_shahar}.
A quantitative estimate for the transition field, depends on the detailed process by which strong phase fluctuations lead to an INS behavior. In this Letter we consider as an example the proliferation of topological excitations as a possible mechanism. We map the
microscopic action of the 2D film onto the 3D XY model. By solving the scaling equations derived for the 3D XY
model~\cite{SRShenoyPRB1995} with corrected initial conditions to
account for both amplitude and phase fluctuations, we get a better agreement with the data of Ref.~\onlinecite{AJohanssonCondMat2006}. We note that the phenomenology of an INS behavior induced by strong phase fluctuations can be generalized to other mechanisms that reduce the phase stiffness in a continuous fashion including magnetic impurities and changing the thickness, as long as dissipation in the cores can be neglected. We would like
to acknowledge useful discussions with E. Altman,  G.
Refael. We thank K.~A.~Parendo, K.~H.~Sarwa, A~Goldman, A.~Johansson and D. Shahar for letting us use their data. This paper was supported by an ISF and a DIP grant.

\bibliographystyle {h-physrev3}

\begin{thebibliography}{10}
\bibitem{CommentDirtyLimit}
The expansion is done in the dirty limit, $\Delta_0\tau\ll 1$, where $\tau$ is
  the scattering mean free time.
\end{thebibliography}

\newcommand{\noopsort}[1]{} \newcommand{\printfirst}[2]{#1}
\newcommand{\singleletter}[1]{#1} \newcommand{\switchargs}[2]{#2#1}
\providecommand{\bysame}{\leavevmode\hbox
to3em{\hrulefill}\thinspace}
\providecommand{\MR}{\relax\ifhmode\unskip\space\fi MR }
\providecommand{\MRhref}[2]{%
  \href{http://www.ams.org/mathscinet-getitem?mr=#1}{#2}
} \providecommand{\href}[2]{#2}

\end{document}